\documentclass[11pt]{article}
\usepackage{graphicx}
\usepackage{amsmath, amssymb, amsfonts}
\usepackage[ansinew]{inputenc}
\usepackage{float}
\newtheorem{theorem}{Theorem}
\newtheorem{proposition}{Proposition}

\newtheorem{remark}{Remark}
\newtheorem{assumption}{Assumption}

\newcommand{\R}{\mathbb{R}}

\usepackage{color}

\def\RR{{\rm I\kern-.17em R}}

\begin{document}
\title{Initial boundary-value problem for the spherically symmetric Einstein equations with fluids with tangential pressure}
\author{Irene Brito \thanks{e-mail: ireneb@math.uminho.pt}, Filipe C. Mena \thanks{e-mail: fmena@math.uminho.pt},\\
{\small $^1$Centro de Matem\'atica, Universidade do Minho, 4710-057 Braga, Portugal.}}
\maketitle

\begin{abstract}
We prove that, for a given spherically symmetric fluid distribution with tangential pressure on an initial spacelike hypersurface with a timelike boundary, there exists a unique, local in time solution to the Einstein equations in a neighbourhood of the boundary. As an application, we consider a particular elastic fluid interior matched to a vacuum exterior.
\\\\
Keywords: Einstein equations; General Relativity; Initial boundary value problems; Self-gravitating systems; Spacetime matching, Elasticity
\end{abstract}

\section{Introduction}

The initial value problem for the Einstein equations for perfect fluids, with suitable equations of state, is well understood in domains where the matter density is positive \cite{Choquet}. However, in physical models of isolated bodies in astrophysics one faces problems where the matter density has compact support and there are matter-vacuum interfaces. From the mathematical point of view, these physical situations can be treated as initial boundary value problems for partial differential equations (PDEs). These cases arise frequently in studies of numerical relativity (see e.g. \cite{Sarbach}) and it is, therefore, important to have analytical results complementing the numerical frameworks.

Rendall \cite{R} proved the existence of local (in time) solutions of an initial value problem for perfect fluid spacetimes with vacuum interfaces. The fluids had polytropic equations of state and vanishing matter density at the interface.
Initial boundary value problems (IBVP), where the matter density vanishes at the interface, were also studied by  Choquet-Bruhat and Friedrich for charged dust matter in \cite{CBF}, where not only existence but also uniqueness of solutions have been proved. In both cases, the field equations were written in hyperbolic form in wave coordinates and no spacetime symmetries were required.

Problems where the matter density does not vanish at the interface bring a non-trivial discontinuity along the boundary. For the Einstein-fluid equations, Kind and Ehlers \cite{KE} use an equation of state for which the pressure vanishes for a positive value of the mass density. They proved that, for a given spherically symmetric perfect fluid distribution on a compact region of a spacelike hypersurface, and for a given boundary pressure, there exists locally in time a unique spacetime that can be matched to a Schwarzschild exterior if and only if the boundary pressure vanishes. In turn, the only existing result along those lines without special symmetry assumptions is due to Andersson, Oliynyk and Schmidt \cite{Andersson} for elastic bodies also having a jump discontinuity in the matter across a vacuum boundary. Under those circunstances, they prove local existence and uniqueness of solutions of the IBVP for the Einstein-elastic fluid system under some technical assumptions on the elasticity tensors. As in all cases above, in \cite{Andersson}, the boundary is characterised by the vanishing of the normal components of the stress-energy tensor although, in that case, conditions on the continuity of the time derivatives of the metric are also imposed. These {\em compatibility conditions} arise naturally from the matching conditions across the matter-vacuum boundary and have to be imposed on the allowed initial data.

The present paper generalises the Elhers-Kind approach to fluids with tangential pressure in spherical symmetry. This includes some cases of interest, such as particular cases of elastic matter. Unlike \cite{KE}, we cannot ensure that, in general, the origin of the coordinates remains regular locally during the evolution. However, this can be ensured in some physically interesting cases.

The plan of the paper is as follows: In Section 2, we setup our IBVP specifying the initial and boundary data. In Section 3, we obtain a first order symmetric hyperbolic (FOSH) system of PDEs and write our main result, which states existence and uniqueness of smooth solutions to the IBVP in a neighbourhood of the boundary. Section 4 contains an application of our results to elastic fluids with vanishing radial pressure and a regular centre.

We use units such that $c=8\pi=1$, greek indices $\alpha, \beta, ..=0,1,2,3$ and latin indices $a,b,..=1,2,3$.

\section{The initial boundary value problem}

Consider a spherically symmetric spacetime $(M,g)$ with a boundary $S$ and containing a fluid source. This gives rise to a fluid 4-velocity $u$ and we define a time coordinate $T$ such that $u$ is normal to the surfaces of constant $T$. We also introduce a comoving radial coordinate $R$.

The general metric for spherically symmetric spacetimes can be written, in comoving spherical coordinates, as \cite{Kramer}
\begin{equation}
\label{metric}
g=-e^{2\Phi(T,R)}dT^2+e^{2\Lambda(T,R)}dR^2+r^2(T,R)d\Omega^2,
\end{equation}
where $d\Omega^2=d\theta^2 +\sin^2\theta d\phi^2,$ and the components of the 4-velocity are written as
\begin{equation}
u^{\mu}=(e^{-\Phi},0,0,0).
\end{equation}
There is freedom in scaling the $T$ and $R$ coordinates which we fix by imposing
\begin{equation}\label{Phir}
\Phi(T,R_0)=0,\;\;\;
 r(0,R)=R,
\end{equation}
where $R_0$ will correspond to the boundary of the matter.
For fluids with no heat flux, the components of the energy-momentum tensor $T_{\mu\nu}$,
in the above coordinates, can be written as
\begin{align}
\label{Tab-spherical}
T_{TT}=\rho e^{2\Phi},\;\;
T_{RR}=p_1 e^{2\Lambda},\;\;
T_{\theta\theta}=p_2 r^2,\;\;
T_{\phi\phi}=p_2 r^2 \sin^2 \theta,
\end{align}
where $\rho$ is the fluid energy density, $p_1$ the {\em radial pressure} and $p_2$ the {\em tangential pressure}.
\begin{assumption} The equation of state for $p_1$ and the energy conditions are such that
\begin{eqnarray}
p_1=p_1(\rho)&\in &C^\infty\label{cond1}\\
\rho&>&0\\
\rho+p_1&>&0\label{cond4}\\
s_{1}^{2}(\rho):=\frac{\text{d} p_{1}(\rho)}{\text{d}\rho}&\ge &0.
\end{eqnarray}
\end{assumption}
\begin{assumption}
The equation of state for $p_2$ is such that
\begin{equation}
\label{pressure2}
p_2=p_2(\rho)\in C^\infty
\end{equation}
and we use the notation
$$
s_{2}(\rho):=\frac{\text{d} p_{2}(\rho)}{\text{d}\rho}.
$$
\end{assumption}
We note that although we do not assume that $s_1^2$ necessarily remains positive when $p_1=0$, as in \cite{KE}, the system of PDEs that we derive for the general case becomes singular for $s_1^2=0$, so we will have to treat this case separately.

\subsection{Einstein and matter equations}

The conservation of the energy-momentum tensor, $\nabla_\nu T^{\mu\nu}=0$, implies
\begin{align}
r\dot{\rho}+ r\dot{\Lambda} (\rho+p_1)+2\dot{r}(\rho+p_2)=0,\;\;\;{\rm for}\;\;\; \mu=T,\label{T1}\\
r\Phi' (\rho +p_1)+rp_1'+2 r' (p_1 -p_2)=0,\;\;\;{\rm for}\;\;\; \mu=R,\label{cT2}
\end{align}
where the prime and dot indicate derivatives with respect to $R$ and $T$, respectively.
The Einstein equations $G_{\mu\nu}=T_{\mu\nu}$ lead to
\\\\
\noindent$(\mu\nu)=(TT)$:
\begin{align}\label{EFE1}
 \rho = \frac{1}{r^2}\left[1-r'^2 e^{-2\Lambda} +\dot{r}^2 e^{-2\Phi}+2r\dot{r}\dot{\Lambda}e^{-2\Phi}-2r(r''-r'\Lambda')e^{-2\Lambda}\right]
\end{align}
$(\mu\nu)=(RR)$:
\begin{align}\label{EFE2}
p_1 =- \frac{1}{r^2}\left[1-r'^2 e^{-2\Lambda} +\dot{r}^2 e^{-2\Phi}-2r r' \Phi' e^{-2\Lambda}+2r(\ddot{r}-\dot{r}\dot{\Phi})e^{-2\Phi}\right]
\end{align}
$(\mu\nu)=(RT)$:
\begin{align}\label{e1}
\dot{r}\Phi'+r' \dot{\Lambda} -\dot{r}'=0
\end{align}
$(\mu\nu)=(\theta\theta)=(\phi\phi)$:
\begin{align}\label{p2}
p_2 &=\frac{e^{-2\Phi}}{r}\left[\dot{r}(\dot{\Phi}-\dot{\Lambda})-\ddot{r}\right]+
e^{-2\Phi}\left[\dot{\Lambda}(\dot{\Phi}-\dot{\Lambda})-\ddot{\Lambda}\right]\nonumber\\
&+\frac{e^{-2\Lambda}}{r}\left[r'(\Phi'-\Lambda')+r''\right]+
e^{-2\Lambda}\left[\Phi' (\Phi'-\Lambda')+\Phi''\right],
\end{align}
while the contracted Bianchi identities are identically satisfied. We note that \eqref{p2} can be obtained from \eqref{T1}-\eqref{e1}.
Integrating (\ref{cT2}), with (\ref{Phir}), one gets
\begin{equation}
\label{phieq}
\Phi(T,R)=-\int_{\rho_0}^{\rho}\frac{s_{1}^{2}(\bar \rho) }{\bar\rho+p_1(\bar\rho)}\; \text{d}\bar\rho- 2 \int_{R_0}^{R}\frac{(p_{1}-p_{2})}{\rho+p_1} \frac{r'}{r}dR.
\end{equation}
\begin{remark}
As an example, in the case of linear equations of state $p_1=\gamma_1 \rho$ and $p_2=\gamma_2 \rho$, we simply get
$$
\Phi(T,R)=-\frac{\gamma_1}{1+\gamma_1}\ln {\left(\frac{\rho}{\rho_0}\right)}-\frac{\gamma_1-\gamma_2}{1+\gamma_1}\ln \left(\frac{r}{r_0}\right)^2,~~{\text with}~~r_0=r(T,R_0),
$$
which will happen for a particular case of elastic matter that we will consider in Section 4.
\end{remark}
Defining the radial velocity as
\begin{equation}\label{rv}
v:=e^{-\Phi}\dot{r}
\end{equation}
and the mean density of the matter within a ball of coordinate radius $R$ as
\begin{equation} \label{md}
\mu:=\frac{3}{r^3}\int_{0}^{R}\rho r^2 r' \text{d}\bar R,
\end{equation}
one obtains from (\ref{EFE1})
\begin{equation}\label{c1}
r'^2 e^{-2\Lambda}=1+v^2 -\frac{1}{3}\mu r^2,
\end{equation}
where we also used the condition $r(T,0)=0$, for regularity of the metric at the center.
Then, (\ref{T1}), (\ref{EFE2}) and (\ref{e1}), together with the evolution equations for $v$ and $\mu$, give
\begin{align}
\dot{r}&=v e^{\Phi},\label{ee1}\\
\dot{v}&=\left[-\frac{r}{2}\left(\frac{\mu}{3}+p_{1}\right)-\frac{r' p_{1}'}{\rho +p_1}e^{-2\Lambda}- 2 \frac{p_1 - p_2}{\rho+p_1}\frac{r'^2}{r}e^{-2\Lambda}\right]e^{\Phi},\label{ee2}\\
\dot{\rho}&=\left[-\rho \left(\frac{v'}{r'}+2\frac{v}{r}\right)-\left(p_1 \frac{v'}{r'}+2p_2 \frac{v}{r}\right)\right]e^{\Phi},\label{ee3}\\
\dot{\Lambda}&=\frac{v'}{r'} e^{\Phi},\label{ee4}\\
\dot{\mu}&=-3\frac{v}{r}(\mu +p_1) e^{\Phi}\label{ee5}.
\end{align}
To summarize, the Einstein equations resulted in the system of evolution equations \eqref{ee1}-\eqref{ee5} for the five variables $r,v,\rho,\Lambda,\mu$ together with constraints \eqref{md} and \eqref{c1}. We note that although we could close the system without \eqref{md} and \eqref{ee5}, those equations will be crucial to obtain a symmetric hyperbolic form. In Section 3, we shall apply suitable changes of variables in order to write our evolution system as FOSH system.

\subsection{Initial data and boundary data}

In spherical symmetry, the free initial data (at $T=0$ and $R\in [0,R_0]$) is expected to be\footnote{We use a "tilde" for boundary data defined on $(T, R=R_0)$ and a "hat" for initial data defined on $(T=0, R)$.} $\hat{\rho}( R):=\rho(0,R)$ and $\hat{v}( R):=v(0,R),$ satisfying $\hat{v}(0)=0$, and constrained by
\begin{equation}
\frac{1}{R}\int_{0}^{R} \hat{\rho}(\bar{R})\bar{R}^2 \text{d}\bar{R}\leq 1 + \hat{v}(R)^2,
\end{equation}
as a consequence of (\ref{Phir}), (\ref{md}) and (\ref{c1}).
The initial data for the remaining variables $r, \mu$ and $\Lambda$ can be obtained from (\ref{Phir}), (\ref{md}) and (\ref{c1}), respectively.

Note that the intrinsic metric and extrinsic curvature (i.e. the first and second fundamental forms) of the initial hypersurface
\begin{equation}
\label{intrinsic}
h_{0}= e^{2\hat{\Lambda}}dR^2+R^2 d\Omega^2,~~~~~
K_0=\hat{v}' e^{2\hat{\Lambda}}dR^2 + R \hat{v}d\Omega^2
\end{equation}
are fully known once $\hat{\rho}( R)$ and $\hat{v}( R)$ are known.

At the boundary of the fluid, we must specify the two smooth boundary functions $\tilde{p}_1(T):=p_1(T,R_0)$ and $\tilde w (T):=v(T,R_0)/r(T,R_0)$ (or $\tilde\mu(T):=\mu(T,R_0)$, via \eqref{ee5}) which should satisfy
the corner conditions $\tilde p_1 (0)=\hat p_1(R_0)$ and $\tilde w (0)=\hat w(R_0)$ (or $\tilde \mu(0)=\hat \mu(R_0)$). In terms of the initial data set $\{\hat \rho(R), \hat v(R)\}$, we note that  from \eqref{ee5} we get
$\left(\dot\mu/(\mu+p_1)\right)(0,R_0)=-3\hat{v}(R_0)/R_0$, at the corner.

The fact that we need $\tilde w(T)$ at the boundary is reminiscent of the compatibility conditions of \cite{Andersson} arising from the matching conditions, since $\tilde w(T)$ is related to the time derivative of the metric at the boundary. In fact, in spherical symmetry, the matching conditions (see the appendix) imply the continuity of the areal radius (here $r$) through the boundary. Therefore, $\dot r(T,R_0)$ also has to be continuous and, for $\Phi(T,R_0)=0$, this gives $\tilde w(T)$, for the interior spacetime from the data of the exterior.

In what follows, we will prove existence and uniqueness of solutions to the evolution equations (\ref{ee1})-(\ref{ee5}), on a neighbourhood of the boundary, for the variables $r,v,\rho, \Lambda,\mu,$ subject to the specified initial and boundary data. We will also show that, since the initial data obeys the constraints (\ref{md}) and (\ref{c1}), the solutions will also satisfy the constraints.
When the fluid boundary corresponds to characteristics, then the corner data will locally determine the boundary evolution and this will be the case when $s_1^2=p_1=0$, as we will show in Section 3.2.
\section{Existence and uniqueness results on a neighbourhood of the boundary}
We treat separately the cases $s^2_1> 0$ and $s^2_1=0$.

\subsection{Case $s_1^2>0$}

In this case, the boundary is non-characteristic and we will be able to use the theorem of Kind-Elhers \cite{KE} provided we write our evolution system as a FOSH system and give the appropriate data. We thus recall the theorem (whose proof uses results of Courant and Lax \cite{CH}):

\begin{theorem} \label{theo-KE}[Kind-Elhers]
Consider the system
\begin{align}
&\dot{X}+A(U_i)Y'=F(X,Y,U_i,R)\nonumber\\
&\dot{Y}+A(U_i)X'=G(X,Y,U_i,R)\label{s1}\\
&\dot{U}_j = H_j (X,Y,U_i,R), \;\; i, j=1,...,p,\nonumber
\end{align}
where $F,G,H_j$ and $A$ are ${C}^{k+1}$ functions, for $R> 0,$ and $A$ is always positive. Let ${C}^{k+1}$ initial values $\hat{X}, \hat{Y},\hat{U_i}$ on $[R_1 ,R_0 ]$, $R_1>0$, and the ${C}^{k+1}$ boundary value $\tilde{Y}(T)$ be given. Assume that $\tilde{Y}(0),\dot{\tilde{Y}}(0),...,{\tilde{Y}}^{(k+1)}(0)$
equal the values of $Y,\dot{Y},...,{Y}^{(k+1)}$ at $(0,R_0),$ which are obtained from \eqref{s1} and the initial data. Then, the system \eqref{s1} has a unique ${C}^k$ solution on a compact trapezoidal domain $\mathcal{T},$ for small enough times.
\end{theorem}
\begin{figure}[H]
\centering
    \includegraphics[width=9cm]{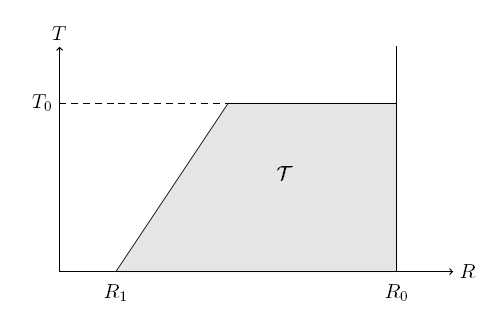}
  \caption{Compact trapezoidal domain $\mathcal{T}$ with small $T_0$.}\label{fig2}
\end{figure}
In order to apply this theorem we will need to use new variables and write the evolution system (\ref{ee1})-(\ref{ee5}) in quasi-linear symmetric hyperbolic form.
We thus use the Kind-Elhers variables
\begin{align}
\mathcal{Q}&=\ln\left(\frac{R}{r}\right),\label{Q}\\
\mathcal{L}&=\int_{\rho_0}^{\rho}\frac{s_1(\rho)}{\rho+p_1(\rho)} \text{d}\rho,\label{L}\\
\omega &=r',\label{o}\\
w&= \frac{v}{r}\label{w},\\
X&=e^{-\Lambda}\mathcal{L}',\label{X}\\
Y&=\frac{v'}{\omega}+2\frac{v}{r}.\label{Y}
\end{align}
Our system will then have the 8 variables
\begin{equation}
\label{variables}
X,~~Y~~{\text{and}}~~U_i=\{{\cal Q}, {\cal L}, \omega, w, \mu, \Lambda\}.
\end{equation}
Before proceeding, note that, from \eqref{ee1}, one obtains
\begin{equation}
\dot{\omega}=(v'+v\Phi')e^{\Phi}.\label{do}
\end{equation}
Now, since \eqref{L} is invertible, we can consider $\rho$ as a known function of ${\cal L}$. Moreover, for given equations of state, we can also consider $p_1, p_2, s_1^ 2$ and $s_2$ as known functions of $\mathcal{L}$. Regarding $\Phi$, it is not clear from \eqref{phieq} that it can be written as a smooth function of the new variables, so we need a further assumption:
\begin{assumption}
$\Phi$, as obtained from \eqref{phieq}, is a known smooth function of the variables $\{X,Y,U_i,R\}$.
\end{assumption}
\begin{remark}
Fulfilling Assumptions 1, 2 and 3 depends on the type of matter and equations of state under consideration. For example, for the linear equations of state of Remark 1 we get:
\begin{eqnarray}
\rho ({\cal L})&=& \rho_0 e^{(1+\gamma_1){\cal L}/\sqrt{\gamma_1}}\\
\Phi({\cal L}, {\cal Q}, R)&=& -\sqrt{\gamma_1}{\cal L}-\frac{2(\gamma_1-\gamma_2)}{1+\gamma_1}\left [\ln {\frac{R}{r_0}}+{\cal Q}\right],
\end{eqnarray}
which for $\rho_0>0, r_0>0$ and $\gamma_1>0$ satisfy the assumptions. As another example, there are cases where the coordinate system can be chosen to be synchronous and comoving so that $\Phi\equiv 0$ and Assumption 3 becomes trivial.
\end{remark}
Then, taking into account the Assumptions 1 and 2, and after a long calculation, our evolution system in terms of the new variables becomes:
\begin{align}
\dot{\mathcal{Q}}&=-w e^{\Phi},\label{e1a}\\
\dot{\mathcal{L}}&=-s_1 Y e^{\Phi}-2s_1 we^{\Phi}\frac{p_2 -p_1}{\rho + p_1},\label{e1b}\\
\dot{\Lambda}&= (Y-2w) e^{\Phi},\label{e1c}\\
\dot{w}&=-e^{\Phi}\left[s_1 \omega e^{\mathcal{Q}-\Lambda}\frac{X}{R}+w^2+\frac{1}{2}\left(\frac{\mu}{3}+p_1 \right)+
2 \frac{\omega^2}{R^2} e^{2(\mathcal{Q}-\Lambda)} \frac{p_1 -p_2}{\rho +p_1} \right],\label{e1d}\\
\dot{\omega}&=e^{\Phi}\left[\omega(Y-2w)-s_1 wXe^{\Lambda-Q}R- 2 \frac{\omega}{R}e^{\mathcal{Q}}\,\frac{p_1 -p_2}{\rho+p_1}\right],\label{e1e}\\
\dot{\mu}&=-3we^{\Phi}(\mu+p_1),\label{e1f}\\
\dot{X}+s_1 e^{\Phi-\Lambda} Y'&=e^{\Phi}\left[XY\left(s_{1}^{2}-\frac{\text{d}s_1}{\text{d}\mathcal{L}}-1\right)+2wX\right]
-2wXe^{\Phi}\frac{p_1-p_2}{\rho+p_1}\nonumber\\
&+e^{\Phi-\Lambda}\frac{p_1-p_2}{\rho+p_1}
\left[2wXe^{\Lambda}\frac{\text{d}s_1}{\text{d}\mathcal{L}}+2 s_1 \frac{\omega}{R}e^{\mathcal{Q}}\left(2Y-3 w\right)-4s_{1}^{2} w Xe^{\Lambda}\right.\nonumber\\
&\left.- 4 s_1 \frac{\omega}{R}we^{\mathcal{Q}}\frac{p_1 -p_2}{\rho +p_1}\right]+2 s_1 w X e^{\Phi}\left(s_1 -\frac{s_{2}}{s_1}\right),\label{e1g}\\
\dot{Y}+s_1 e^{\Phi-\Lambda} X'&=e^{\Phi}\left[X^2\left(s_{1}^{2}-\frac{\text{d}s_1}{\text{d}\mathcal{L}}\right)-2s_1 \omega e^{Q-\Lambda}\frac{X}{R}-(Y-2w)^2-2w^2-\frac{\rho+3p_1}{2}\right]\nonumber\\
&+2\frac{X}{R} e^{\Phi-\Lambda+\mathcal{Q}}\left(\frac{s_{2}}{s_1}-s_1\right)+ e^{\Phi}\frac{p_1 -p_2}{\rho +p_1}\left[\left(6s_1 +\frac{2}{s_1}\right)\omega \frac{X}{R}e^{\mathcal{Q}-\Lambda}-3\mu+p_1\right.\nonumber\\
&\left.+2\rho- 2\frac{\omega^2}{R^2}e^{2(\mathcal{Q}-\Lambda)}\left(1-2\frac{p_1 -p_2}{\rho +p_1}\right)+ 2 Yw\left(2 w-1\right)\right],\label{e1k}
\end{align}
which has the symmetric hyperbolic form \eqref{s1}, under Assumption 3. Note that the system reduces to the one of \cite{KE} for $p_1\equiv p_2$.

In our system, the equation \eqref{e1a} was obtained from \eqref{Q} using \eqref{ee1}, while \eqref{e1b} came from \eqref{L} together with \eqref{ee3}. The evolution equation for $\Lambda$ was derived from \eqref{ee4}, and the evolution equation for $w$ from \eqref{w} together with \eqref{ee2}. Equation \eqref{e1e} came from \eqref{do}. The evolution equations for $\mu, X$ and $Y$ were obtained from \eqref{ee5}, \eqref{X} and \eqref{Y}, respectively.

The constraints are given by the equations \eqref{o}, \eqref{X}, \eqref{Y} and by the derivatives of \eqref{md} and \eqref{c1} with respect to $R$. They can be expressed in the following way
\begin{align}
C_1&:= \mathcal{L}'-e^{\Lambda}X=0,\\
C_2&:=R\mathcal{Q}'+\omega e^{Q}-1=0,\\
C_3&:= Re^{-Q}w'-\omega(Y-3w)=0,\\
C_4&:= Re^{-Q}\mu'+3\omega (\mu-\rho)=0,\\
C_5&:=e^{-2\Lambda}(\omega'-\Lambda \omega)+Re^{-Q}\left[\frac{1}{6}(3\rho-\mu)-w(Y-2w)\right]=0,
\end{align}
and, as in \cite{KE}, it can be shown that for a given $C^1$ solution $\{X,Y,U_i\}$ of \eqref{e1a}-\eqref{e1f}, the quantities $C_1,..,C_5$ satisfy a linear system of the form
\begin{equation}
\dot{C}_k=\sum_{l=1}^{5}A_{kl}C_l,
\end{equation}
where $A_{kl}$ are continuous functions of $X,Y,U_i$. We then conclude that the constraints $C_k =0$ are satisfied for all $T$ if they are satisfied at $T=0.$

From the initial data $\hat{\rho}(R)$ and $\hat{v}(R),$ using \eqref{Phir}, one obtains
\begin{equation}
\mathcal{L}(0,R)=\mathcal{L}(\hat{\rho}(R)),\;\; w(0,R)=\frac{\hat{v}(R)}{R},\;\; \mathcal{Q}(0,R)=0,
\end{equation}
and the initial data for $\mu,\Lambda,\omega,X,Y$ are specified by \eqref{md}, \eqref{c1}, \eqref{o}, \eqref{X}, \eqref{Y}, respectively. We impose that the quantities $\mu,\Lambda,\omega,X,Y$: (i) satisfy the contraints initially, i.e. $C_k (0,R)=0$, and (ii) are smooth functions in a region $[R_1,R_0]$, for some $R_1>0$.

To apply Theorem \ref{theo-KE}, we also need the necessary boundary function which we get from \eqref{e1b}, considering $\Phi(T,R_0)=0$, as:
\begin{equation}
\tilde{Y}(T)=Y(T,R_0)=-\left[\frac{\dot{\mathcal{L}}}{s_{1}(\mathcal{L})}\right](T,R_0)
-2w(T,R_0)\left[\frac{p_2 (\mathcal{L})-p_1 (\mathcal{L})}{\rho(\mathcal{L})-p_1 (\mathcal{L})}\right](T,R_0),
\end{equation}
which, using \eqref{e1f}, can be rewritten as
\begin{equation}
\label{boundary-Y}
\tilde{Y}(T)=-\left[\frac{\dot{\mathcal{L}}}{s_{1}(\mathcal{L})} +\frac{2}{3}\left(\frac{\dot\mu}{\mu+p_1(\mathcal{L})}\right)\frac{p_2(\mathcal{L})-p_1(\mathcal{L}) }{\rho(\mathcal{L})-p_1(\mathcal{L})}\right](T,R_0).
\end{equation}
This function will be known given the two smooth boundary functions $\tilde{p}_1(T)$ and $\tilde w (T)$ (or $\tilde\mu(T)$) which should satisfy
the corner conditions, as described in Section 2.2.

We are now in the position of applying Kind-Elhers' theorem to \eqref{e1a}-\eqref{e1k}, given the above initial and boundary data,
and this proves existence and uniqueness of solutions to the initial boundary value problem in a neighbourhood of the boundary. In detail, we get the following result:
\begin{theorem}
\label{theo2} Consider a fluid matter field satisfying Assumptions 1, 2 and 3 with $s_1^2>0$.
Then, the system \eqref{e1a}-\eqref{e1k} is a FOSH system of the form \eqref{s1} for the variables $X,Y$ and $U_j=\{\mu,Q, \omega, w, \Lambda,{\cal L}\}$. Suppose that the initial data for those variables is smooth on $[R_1,R_0]$, for some $R_1>0$, and that the constraints $C_1,..,C_5$ are satisfied at $T=0$. Suppose a spherically symmetric distribution of such matter is given together with the smooth boundary functions $\tilde p_1(T)$ and $\tilde w(T)$ (or $\tilde \mu(T)$). Suppose that the given initial and boundary data for $Y$ and their time derivatives satisfy the corner conditions at $(0,R_0)$. Then, there exists locally in time a unique smooth solution to the Einstein equations in ${\cal T}$.
\end{theorem}
A well known case is that of a Kottler spacetime exterior \eqref{SCHmetric} which can be attached to the fluid if and only if $\tilde p_1(T)=-\Lambda$, where $\Lambda\in \R$. A way to see this using our framework is as follows: The conditions of the continuity of the first and second fundamental forms across the boundary (see the appendix) imply the continuity of the normal component of the pressure $\tilde p_1(T)=-\Lambda$ (this is also a well known consequence of the so-called Israel conditions). Due to the Einstein equations and the continuity of the areal radius, this condition turns out to be equivalent to the continuity of the mass through the boundary $\tilde \mu (T)\tilde r^3(T)= 6m,$ where $m$ is the Kottler mass.
%
\subsection{Case $p_1=s_1^2=0$}

While the constraints equations $C_2,..,C_5$ remain the same in the case $p_1=s_1^2=0$,
the quasi-linear system \eqref{e1a}-\eqref{e1f} is modified since $X\equiv {\cal L}\equiv 0$.
In that case, we get a semi-linear symmetric hyperbolic PDE system of the form
\begin{equation}
\label{p1zero}
\dot{U}_j = H_j (U_i,R),~~~ i,j=1,..,6,
\end{equation}
for the variables $U_j=\{Y,\omega, w, {\cal Q}, \mu, \Lambda\}$. An important aspect of this system, compared to the previous one, is that the quantity $s_2$ does not appear in $H_j$. This means that we do not need to use Assumptions 1 and 2 in order to close the system, as before. Instead, we make the following assumption:
\begin{assumption}
Both $\rho>0$ and $p_2$ are smooth known functions of  the variables $\{Y,\omega, w, {\cal Q}, \mu, \Lambda, R\}$.
\end{assumption}
In this case, the characteristics of the system \eqref{p1zero} are the vertical lines of constant $R$ and the trapezoidal region ${\cal T}$ of Figure 1 is now a rectangle. In particular, the boundary is now characteristic. IBVP with characteristic boundaries were investigated, in more generality, by Chen \cite{Chen} and Secchi \cite{Secchi} for quasi-linear systems.

In our case, the integration along the characteristics gives, at each point $(T,  R)$, simply
\begin{equation}
\label{}
U_j(T, R) = U_j (0,R)+\int_0^{T} H_j (U_i, R) d\bar T.
\end{equation}
which, using the methods of Courant and Lax \cite{CH, KE}, gives a smooth local (in time) solution to the system \eqref{p1zero}, once smooth initial data $\hat U_j (R)$ is prescribed. In our case, provided suitable data from an exterior spacetime at the corner $(0,R_0)$, one can integrate from that point to get uniquely the boundary functions $\tilde U_j(T):=U_j(T,R_0)$.
In this sense, the spacetime boundary is completely fixed by the initial corner conditions.

As mentioned in Section 2.2, from the initial data $\{\hat \rho, \hat v\}$ we can get $h_0$ and $K_0$ from the expressions \eqref{intrinsic}. The corner conditions are the matching conditions evaluated at $(0,R_0)$, and these give the components of $h_0$ and $K_0$ at $R_0$.
In particular, this provides the corner data $\hat\Lambda_0:=\hat \Lambda(R_0), \hat v_0:= \hat v(R_0)$ and $\hat v'_0:=\hat v'(R_0)$.  In turn, this gives the remaining corner data for the system
\begin{equation}
\label{corner-data}
\hat\omega(R_0)=\frac{\hat v_0}{R_0},~~ \hat Y(R_0)= \hat v'_0+2\frac{\hat v_0}{R_0},~~ \hat Q(R_0)=0,~~ \hat w(R_0)=1,~~ \hat \mu(R_0)= -\frac{3}{R_0^2} (e^{-2\hat\Lambda_0}-1-\hat v^2_0).
\end{equation}
We summarize this discussion as:
\begin{theorem}
\label{prop1}
Consider a matter field with $p_1=s_1^2=0$ and satisfying Assumptions 3 and 4.
Then, the system \eqref{e1a}-\eqref{e1k} reduces to the form \eqref{p1zero} which corresponds to a semi-linear FOSH system for the variables $U_j=\{Y,\omega, w, {\cal Q}, \mu, \Lambda\}$. Consider the smooth initial data $\{\hat\rho, \hat v \}$ for the variables $U_j$ on $[R_1,R_0]$, for some $R_1>0$, and suppose that the matching conditions to an exterior spacetime are satisfied at $(0,R_0)$. Suppose that the constraints $C_2,..,C_5$ are satisfied at $T=0$. Then, there exists a unique smooth solution to the Einstein equations in the rectangle $[R_1,R_0]\times T_0$ for small enough $T_0>0$.
\end{theorem}
\section{An application to self-gravitating elastic bodies}

In this section, we consider a simple application of the previous result to elastic matter. In this case, the matter does not necessarily satisfy \eqref{cond1} or \eqref{pressure2}. So, before proceeding, we recall some basic facts about elastic matter adapting the presentation to the particular setting we shall consider in the context of spherical symmetry.

The material space $X$ for elastic matter is a three-dimensional manifold with Riemannian metric $\gamma$.
Points in $X$ correspond to particles of the material and the material metric $\gamma$ measures the distance between particles in the relaxed (or unstrained) state of the material. Coordinates in $X$ are denoted by $y^A$, $A=1,2,3.$ The spacetime configuration of the material is described by the map $\psi: M \longrightarrow X$, where $M$ denotes the spacetime with metric $g$.
The differential map $\psi_{\ast}: T_{p}M\longrightarrow T_{\psi(p)}X$, which is also called relativistic deformation gradient, is represented by a rank 3 matrix with entries $\left(y^{A}_{\mu}\right)=\frac{\partial y^A}{\partial x^{\mu}}$, where $x^{\mu}$ are coordinates in $M$. In turn, the matrix kernel is generated by the 4-velocity vector $u$ satisfying $y^{A}_{\mu}u^{\mu}=0.$

The push-forward of the contravariant spacetime metric from $M$ to $X$ is defined by
\begin{equation}
G^{AB}= \psi_{\ast}g^{\mu\nu}=g^{\mu\nu}y^{A}_{\mu}y^{B}_{\nu},
\end{equation}
which is symmetric positive definite and, therefore, a Riemannian metric on $X$. This metric contains information about the state of strain of the material, which can be described by comparing this metric with structures in $X$, e.g. the material metric. The material is said to be unstrained at an event $p\in M$ if $G_{AB}=\gamma_{AB}$ at $p.$

The dynamical equations for the material can be derived from the Lagrangian density $L=-\sqrt{-\text{det}g}\rho,$
where
\begin{equation}\label{rho}
\rho=\epsilon h
\end{equation}
 is the rest frame energy per unit volume. The density of the matter, measured in the material rest frame, is given by
\begin{equation}
\epsilon=\tilde{\epsilon}(y^a) \sqrt{\text{det}G^{AB}},
 \end{equation}
 where $\tilde{\epsilon}(y^a)$ is an arbitrary positive function.
 The equation of state is defined by the function $h=h(y^A, G^{AB}),$ which describes the dependence of the energy on the state of strain and specifies the material.

The stress-energy tensor for elastic matter can be written as \cite{MagliI}
\begin{equation}
T^{\mu}_{~\nu}=\epsilon\left(-\frac{h}{g_{00}}\delta_{~0}^{\mu}g_{0 \nu}+2\frac{\partial h}{\partial G^{AB}}\delta_{~\nu}^{A}g^{\mu B}\right).
\end{equation}
In spherical symmetry, the push-forward of the metric $g$ with line-element \eqref{metric} gives
\begin{equation}
G^{AB}=\eta\delta^{A}_{~1}\delta^{B}_{~1}+\beta\delta^{A}_{~2}\delta^{B}_{~2}
+\tilde{\beta}\delta^{A}_{~3}\delta^{B}_{~3},
\end{equation}
where $\eta=e^{-2\Lambda},$  $\beta=1/r^2$ and $\tilde{\beta}=\beta/\sin^2{\theta}$.
In the unstrained state, one has $\Lambda=0$ and $r=R,$ so that $\eta=1$ and $\beta=1/R^2.$

The matter density of the material is, in our case, given by
\begin{equation}
\epsilon=\epsilon_{0}(R)\beta\sqrt{\eta},
\end{equation}
assuming that $\tilde{\epsilon}(y^A)=\epsilon_{0}(R)\sin\theta,$ where $\epsilon_{0}$ is an arbitrary positive function.
The energy-momentum tensor can then be expressed by
\begin{equation}
T^{\mu}_{~\nu}=\text{diag}(-\rho,p_1 ,p_2 ,p_2),
\end{equation}
where $\rho=\epsilon h,$ $p_1 = 2\epsilon \eta \frac{\partial h}{\partial \eta},$ $p_2 = -\frac{1}{2}\epsilon r \frac{\partial h}{\partial r}$ and $h=h(R,\eta,r)$. Using \eqref{rho}, we get
\begin{equation}
p_1 = 2\frac{\eta}{h}  \frac{\partial h}{\partial \eta}\rho, ~~~~~p_2 = -\frac{1}{2}\frac{r}{h} \frac{\partial h}{\partial r}\rho.\label{p2M}
\end{equation}
This $T^{\mu}_{~\nu}$ for elastic matter falls in the class given by \eqref{Tab-spherical} although, in general, we will not have $p_1=p_1(\rho)$ as in Assumption 1, so Theorem 2 does not apply. We will now investigate a particular elastic fluid for which $p_1=s_1^ 2=0$ and, in that case, we may use Theorem 3.
\subsection{Magli's ansatz}
Magli \cite{MagliI} found a class of non-static spherically symmetric solutions of the Einstein equations corresponding to anisotropic elastic spheres. These models have vanishing radial stresses and generalize the Lema\^itre-Tolman-Bondi dust models of gravitational collapse, by including tangential stresses.

Assuming that the equation of state for the elastic matter, prescribed by the function $h$, does not depend on the eigenvalue $\eta$ in the material space, i.e. $h=h(R,r),$ then it is possible to write the spherically symmetric metric as \cite{MagliII}
\begin{equation}
\label{metric-magli}
ds^2=-e^{2\Phi}dt^2 +\frac{r'^2 h^2}{1+f}dR^2+r^2d\Omega^2,
\end{equation}
where
\begin{equation}\label{eqPhi}
\Phi (T,R)=-\int \frac{1}{h}\frac{\partial h}{\partial r}r' \text{d}R+c(T),
\end{equation}
and $c(T)$ is an arbitrary function, reflecting the invariance with respect to time rescaling, which can be chosen such that $\Phi(T,R_0)=0$, for some $R_0>0$. One can then obtain a first integral to the Einstein equations as
\begin{equation}
\label{dotr}
e^{-2\Phi}\dot{r}^2=-1+\frac{2F}{r}+\frac{1+f}{h^2},
\end{equation}
where $F(R)>0$ and $f(R)>-1$ are arbitrary functions which we assume to be smooth on $[0,R_0]$. With the assumption $h=h(R,r)$, and using the first integral, we also get
\begin{align}
\rho (T,R)&=\frac{F'}{4\pi r^2 r'},\nonumber\\
p_1 (T,R) & = 0, \\
p_2 (T,R)&=-\frac{r}{2h}\frac{\partial h}{\partial r} \rho :=\mathcal{H}\rho, \nonumber
\end{align}
where $\mathcal{H}=-\frac{r}{2h}\frac{\partial h}{\partial r}$ is called adiabatic index (see \cite{MagliII}, cf. \eqref{p2M}).

A physically interesting example of a spacetime with $h=h(R,r)$ is the non-static Einstein cluster in spherical symmetry, which describes a gravitational system of particles sustained only by tangential stresses \cite{Datta, Bondi,MagliI}.

To make contact with our formalism, in this case, we get a semi-linear system of the form \eqref{p1zero},
for the variables $U_j=\{Y,\mathcal{Q},\Lambda, w, \omega, \mu\}$ with smooth functions
\begin{align}
\rho ({\cal Q}, \omega, R)&= \frac{F'(R) e^{\cal Q}}{4\pi \omega R}\\
p_2  ({\cal Q}, \omega, R) &= \mathcal{H}({\cal Q}, R) \rho ({\cal Q}, \omega, R),
\end{align}
which clearly satisfy Assumption 4,
and smooth  initial data on $[R_1,R_0]$, $R_1>0$, as
\begin{align}
\hat{Y}&=\hat{\Omega}'+2\frac{\hat{\Omega}}{R},\;\; \text{where}\;\; \hat{\Omega}=\sqrt{\frac{1+f}{\hat{h}^2}-\left(1-\frac{2F}{R}\right)},\nonumber\\
\hat{\mathcal{Q}}&=0,\nonumber\\
\hat{\Lambda}&=\frac{1}{2}\ln\left(\frac{\hat{h}^2}{1+f}\right),\nonumber\\
\hat{w} &= \frac{1}{R}\sqrt{\frac{1+f}{\hat{h}^2}-\left(1-\frac{2F}{R}\right)},\nonumber\\
\hat{\omega} &=1,\nonumber\\
\hat{\mu}&=\frac{3(F(R)-F(0))}{4\pi R^3}.\nonumber
\end{align}
The free initial data here is given by a smooth function $\hat h> 0$, i.e. the initial equation of state, and smooth functions $F$ and $f$ which come from the initial density $\hat \rho$ and radial velocity $\hat v$ profiles, respectively, as in Section 2.2. For smooth $\hat h, F, f$ on some interval $[R_1,R_0]$, $R_1>0$, we are under the conditions of Theorem 3 which ensures existence and uniqueness of solutions in a neighbourhood of the boundary.

In this case, we can also obtain similar results in a neighbourhood of the origin $R=0$, at least for some choices of initial data. This has to be done separately since our evolution system \eqref{p1zero} is singular at the center.
\begin{figure}[H]
\centering
    \includegraphics[width=9cm]{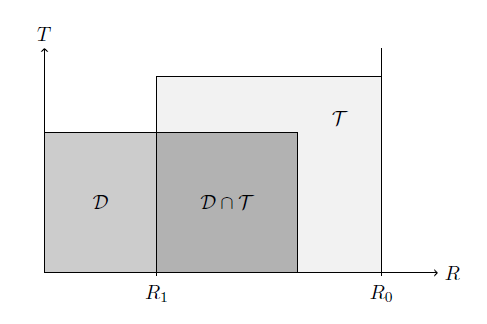}
  \caption{Domains ${\cal D}$ and ${\cal T}$, where in ${\cal D} \cap {\cal T}$ the solutions agree due to uniqueness.}\label{fig2}
\end{figure}
Consider an open region ${\cal D}$ including the center, as in Figure \ref{fig2}, by considering the original variables of Magli. In that case, the center is regular for $r(T,0)=0$, smooth functions $f$ and $F$ such that \cite{MagliI}
\begin{equation}
f(0)=h^2(0,0)-1~~~{\text{and}}~~~F(R)=R^3\varphi(R),~~{\text{with}}~~\varphi(0)~~{\text{is finite}},
\end{equation}
and equations of state satisfying the {\em minimal stability requirement}, namely that $h$ has a minimum at $r=R$ \cite{MagliI}.
Physically, this means that the centre is unstrained and has zero radial velocity as $\lim_{R\to 0} \dot r=0$.

In \cite{MagliI}, Magli shows that there are open sets of $C^\infty$ data $\hat h, F$ and $f$ such that the regularity conditions at the centre are fulfilled.
From the PDE point of view, this implies that for such data there exists a unique $C^\infty$ solution to \eqref{dotr}, for small enough time in a neighbourhood ${\cal D}$ of the centre. Uniqueness of the spherically symmetric initial data gives the uniqueness of solutions in the region ${\cal D} \cap {\cal T}$.
We have then proved:
\begin{proposition}
	Consider a spacetime with metric \eqref{metric-magli} and containing elastic matter satisfying \eqref{eqPhi}. Consider smooth initial data $F,f,\hat h$ on $[0,R_0]$ satisfying the (corner) matching conditions to an exterior spacetime at $(0,R_0)$ and the regularity conditions at the center. Then, there exists a unique smooth solution to the Einstein equations
on $[0, R_0]\times T_0$, for small enough $T_0$.
\end{proposition}
A suitable exterior to such elastic spacetimes is given by the Schwarzschild solution, which can be matched initially at $R_0$. We omit the details of the matching conditions here since they were given in \cite{MagliI}.

As a final remark, we note that the analysis of this particular IBVP can be taken much further if one has explicit solutions for $r$.
These solutions can be obtained by assuming $h=h(r)>0$. In this case, equations \eqref{eqPhi} and \eqref{dotr} decouple and
$\Phi'=-\frac{1}{h}\frac{d h}{dr}r',$ yielding
$\Phi (T,R)=-\ln h+\ln{ h(r(T,R_0))}$,
which satisfies $\Phi(T,R_0)=0.$ Consider a linear stress-strain relation $p_2 = k \rho,$ where the adiabatic index $k$ satisfies $-1\leq k\leq 1$ in order to comply with the weak energy condition \cite{SW}. This choice of stress-strain is equivalent to set
$h(r)=r^{-2k}$. In that case, one has $\dot{r}^2=2F r^{4k-1}+(1+f)r^{8k}-r^{4k}$ which is now integrable. Remarkably, for some values of $k$, such as $k=-1/4$ and $k=1/4$, one can obtain explicit solutions which satisfy $r(0,R)=R$ and $r(T,0)=0$ and have a regular origin for certain choices of initial data \cite{SW}. This allows e.g. to solve the matching conditions and obtain the boundary hypersurface \cite{MagliI}, as well as to study of global properties of the matched spacetime such as the formation of horizons and spacetime singularities \cite{SW,MagliII}.
\\\\
{\bf Funding:} This work was supported by CMAT, Univ. Minho, through FEDER Funds COMPETE and FCT Projects Est-OE/MAT/UI0013/2014 and PTDC/MAT-ANA/1275/2014.
\\\\
{\bf Acknowledgments:} We thank Robert Beig and Marc Mars for useful comments. We thank the referees for the comments and suggestions.

\section*{Appendix: Matching conditions in spherical symmetry}

In this appendix, for completeness, we recall some basic facts about spacetime matching theory and write the matching conditions in spherical symmetry adapted to our context. These conditions are known but are of importance here since they provide compatibility conditions for our IBVP problem.

Let $(M^{\pm},g^{\pm})$ be spacetimes with non-null boundaries $S^{\pm}$.
Matching them requires an identification of the boundaries, i.e. a pair of embeddings
$\Omega^\pm:\; S \longrightarrow M^\pm$ with $\Omega^\pm(S) = S^{\pm}$,
where $S$ is an abstract copy of any of the boundaries.
Let $\xi^i$ be a coordinate system on $S$. Tangent vectors to
$S^{\pm}$ are obtained by $f^{\pm \alpha}_i =
\frac{\partial \Omega_\pm^\alpha}{\partial \xi^i}$ though we shall work with orthonormal combinations $e^{\pm \alpha}_i$ of the $f^{\pm \alpha}_i$. There are also  unique (up to
orientation) unit normal vectors $n_{\pm}^{\alpha}$ to the boundaries.
We choose them so that if $n_{+}^{\alpha}$ points into $M^{+}$ then
$n_{-}^{\alpha}$ points out of $M^{-}$ or viceversa.
The first and second fundamental forms on $S^\pm$
are simply
$$q_{ij}^{\pm}= e^{\pm \alpha}_i e^{\pm \beta}_j
g_{\alpha\beta}|_{_{S^\pm}},~~
K_{ij}^{\pm}=-n^{\pm}_{\alpha} e^{\pm
\beta}_i\nabla^\pm_\beta e^{\pm \alpha}_j.$$
The matching conditions (in the absence of shells), between two spacetimes $(M^\pm,g^\pm)$ across a non-null hypersurface $S$,
are the equality of the first and second fundamental forms (see \cite{Mars-Seno}):
\begin{equation}
q_{ij}^{+}=q_{ij}^{-},~~~
K_{ij}^{+}=K_{ij}^{-}.
\label{eq:backmc}
\end{equation}
We shall now specify the matching conditions for the case where the interior  has the anisotropic fluid metric \eqref{metric} and the exterior  is the Kottler spacetime given by metric:
\begin{equation}
\label{SCHmetric}
g^{+} = - \left ( 1 - \frac{2m}{\varrho}-\frac{\Lambda}{3}\varrho^2 \right ) dt^2 +
\frac{d\rho^2}{1 - \frac{2m}{\varrho}-\frac{\Lambda}{3}\varrho^2 } +\varrho^2 d\Omega^2.
\end{equation}
The boundary $S^+$ can be parametrized as $\Omega^+=\{ t = t_0 (\lambda), \varrho = \varrho_0 (\lambda) \}$, where two dimensions were omitted, and we can assume $\dot t_0 >0$ without loss of generality. 

Considering our choice $\Phi(T,R_0)=0$ for the interior spacetime, we can take the following parametrization $\Omega^-=\{ T = \lambda, R = R_0\},$ so that $\dot{T}=1$ and $\dot{R}=0.$ Then, the equality of the first fundamental forms on $S$ gives
\begin{align}
\label{matchcond1}
1&= \left ( 1 - \frac{2m}{\varrho_0}-\frac{\Lambda}{3}\varrho_0^2  \right ) \dot t_0^2
- \frac{\dot\varrho_0^2}{1 - \frac{2m}{\varrho_0}-\frac{\Lambda}{3}\varrho_0^2 }, \\
\label{matchcond2}
r(\lambda,R_0)&=\varrho_0(\lambda)
\end{align}
and the equality of the second fundamental forms, on $S$, implies
\begin{align}
\label{matchcond3}
\Phi' e^{2\Phi -\Lambda}&=\left (-\dot t_0 \ddot \varrho_0 + \ddot t_0  \dot\varrho_0+ \frac{3m \dot\varrho_0^2 \dot t_0}{\varrho_0 \left (\varrho_0
- 2m-\frac{\Lambda}{3}\varrho_0^3 \right )} - \frac{m}{\varrho_0^2} \left ( 1- \frac{2m}{\varrho_0}-\frac{\Lambda}{3}\varrho_0^2  \right ) \dot t_0^3
\right ), \\
\label{matchcond4}
 rr'e^{-\Lambda}&=-\dot t_0 \left (\varrho_0  -2 m -\frac{\Lambda}{3}\varrho_0^3 \right ).
\end{align}
Then, the matching conditions give expressions for $\varrho_0 (\lambda)$, $t_0(\lambda)$ and imply the continuity of the mass $\tilde\mu \tilde r^3=6m$ through the boundary, as is well known. By further substituting the Einstein equations, those conditions also imply that $\tilde p_1=-\Lambda$.
In fact, from $\tilde p_1=-\Lambda$ we also recover $\tilde\mu \tilde r^3=6m$ as mentioned at the end of Section 3.1.

Generalising this procedure, consider now the matching where the interior is still the anisotropic fluid \eqref{metric} and the exterior is the spacetime given by a general spherically symmetric metric:
\begin{equation}
\label{sphsymmetric}
g^{+} = - e^{2\mu(t,\varrho)} dt^2 +
e^{2\nu(t,\varrho)}d\varrho^2 +\bar{r}^2(t,\varrho) d\Omega^2.
\end{equation}
The boundaries $S^\pm$ are parametrized as before.
Then, the equality of the first fundamental forms, for $\Phi(T,R_0)=0$, imply
\begin{align}
-1&=-\dot{t}_0^2e^{2\mu}+\dot{\varrho}_0^2 e^{2\nu}, \\
\label{r-equation}
r(\lambda,R_0)&=\bar{r}(t_0(\lambda),\varrho_0(\lambda))
\end{align}
and the equality of the second fundamental forms gives
\begin{align}
-\Phi' e^{-\Lambda}&=e^{\mu+\nu} \left (\dot{\rho}\dot{t}_0^2 \dot \mu +2\dot{\rho}_0^2\dot{t}_0\mu'+\dot{\rho}_0^3\dot{\nu}e^{2\nu-2\mu}-\dot{t}_0^3\mu'e^{2\mu-2\nu}-2\dot{t}_0^2\dot{\rho}_0\dot{\nu}-\dot{\rho}_0^2\dot{t}_0\nu'
\right ), \\
 rr'e^{-\Lambda}&=e^{\mu+\nu}\bar{r}\left( \dot{t}_0e^{-2\nu}\bar{r}'+\dot{\rho}_0e^{-2\mu}\dot{\bar{r}}\right).
\end{align}
The above equations provide $\rho_0(\lambda)$, $t_0(\lambda)$ and $p_1^-(\lambda)=p_1^+(\lambda)$, at the boundary. Another way to get this last relation is to use the Israel conditions $n^\mu_- T^-_{\mu\nu}=n^\mu_+ T^+_{\mu\nu}$ at $S$ (see e.g. \cite{Mars-Seno}). With this boundary data, and knowing the exterior spacetime, one gets $\tilde r(\lambda):=r(\lambda,R_0)$ from \eqref{r-equation} and, therefore, $\tilde w(\lambda)$ as mentioned in Section 3.1.
%

\end{document}